\documentclass[showpacs,showkeys,preprintnumbers,amsmath,amssymb,times,graphicx]{revtex4}
\usepackage{graphicx} 
\usepackage{dcolumn}% Align table columns on decimal point
\usepackage{bm}% bold math

\usepackage{color}
\usepackage{amsmath}
\usepackage{amsfonts}
\usepackage{amssymb}
\usepackage{amsbsy}
\usepackage[figuresright]{rotating}
\usepackage[font=small,labelfont=bf]{caption}
\usepackage{natbib}

\def\3{\ss}

\def\q0{\phantom{1}}

\def\thetabn{\Theta_{Bn}}

\def\parp2{\frac{\partial^{2}}{\partial p^{2} }}

\def\m21{$2^{\circ}\times 1^{\circ}$}
\def\ts{\thinspace}

\def\ne8{Ne\ts{$\scriptstyle {\rm VIII}$} }

\newcommand{\aap}{{Astron.\ Astrophys. }}

\newcommand{\asr}{{Adv. \ Space. \ Res.}}

\newcommand{\ag}{{Ann.\ Geo\-phy\-si\-c\ae}}

\newcommand{\grl}{{Geo\-phys.\ Res.\ Lett.}}

\newcommand{\jgr}{{J.\ Geo\-phys.\ Res.}}

\newcommand{\ssr}{{Space Sci.\,Rev.}}

\newcommand{\pf}{{Phys. Fluids}}

\newcommand{\pop}{{Phys.\ Plasmas}}
\def\ion[#1 #2]{#1\,{\sc #2}}
\def\lamb[#1]{#1\,{\AA}}
\def\serts89{SERTS-89}
\def\fe12{Fe\,{\sc xii}}
\def\mb[#1]{\makebox[0.15cm][l]{#1}}
\begin{document}

\title{The Heliospheric Termination Shock}

\author{R. A. Treumann$^\dag$ and C. H. Jaroschek$^{*}$}\email{treumann@issibern.ch}
\affiliation{$^\dag$ Department of Geophysics and Environmental Sciences, Munich University, D-80333 Munich, Germany  \\ 
Department of Physics and Astronomy, Dartmouth College, Hanover, 03755 NH, USA \\ 
$^{*}$Department Earth \& Planetary Science, University of Tokyo, Tokyo, Japan
}%

\begin{abstract} The heliospheric Termination Shock is the largest (by dimension) shock in the heliosphere. It is believed that it is also the strongest shock and is responsible for the generation of the Anomalous Cosmic Ray component in the heliosphere. This chapter review the gross properties and observations of the Termination Shock. It is structured as follows: 1. The heliosphere, providing the heliospheric stage for Termination Shock formation, 2. The argument for a heliospheric Termination Shock, 3. The global heliospheric system, 4. Termination Shock properties, 5. Observations: the Voyager passages, radio observations, plasma waves and electron beams, traces of plasma and magnetic field, energetic particles, galactic cosmic rays, Termination Shock particles, the anomalous cosmic ray component, 6. Conclusions.
\end{abstract}
\pacs{}
\keywords{}
\maketitle 

\section{The Outer Heliosphere}\label{intro-1}
\noindent
On its way out from the sun into the near interstellar environment of our solar system the super-magnetosonic solar wind is not remarkably retarded by its interaction with the celestial bodies in the solar system. These bodies are not very frequent and, even though large by human measures, retarding the solar wind locally, diverting it locally from its straight radial path and forcing it to flow around their magnetospheres or atmospheres, are too small to affect the solar wind on the global scale. The diverted wind, consisting of charged particles that are slower than the solar wind while being heated by the interaction with the bow shocks, behave like pick-up ions which in their proper system feel the solar wind electric field, become accelerated and couple to the solar wind. In this way the celestial bodies in the solar system create hot plasma striations in the solar wind on the scales of the magnetospheric cross sections which interact with the solar wind by plasma instabilities and gradually heat it when moving down the way to the boundaries of the heliosphere. By this they contribute to the downstream solar wind turbulence and at the same time a little retard the adiabatic cooling of the solar wind which should be the consequence of its three-dimensional collisionless outward expansion. 

It is only in the outer heliosphere that the solar wind becomes retarded mainly by its interaction with the interstellar gas (conventionally  called the Local Interstellar Medium LISM) which it encounters with its full bulk kinetic energy density ${\cal E}_{\rm SW}(r)=\frac{1}{2}m_iN(r)V_1^2$. Since the solar wind density decreases approximately proportional to the surface of a sphere, the bulk flow energy density is roughly given by ${\cal E}_{\rm SW}(r)={\cal E}_{{\rm SW},1}R^{-2}$, where $R=r/(1\,{\rm AU})$ is the heliocentric radial distance measured in AU, and ${\cal E}_{{\rm SW},1}\approx 10^7\,{\rm keV/m^3}\approx 10^{-9}\,{\rm J/m^3}$ is the average bulk solar wind energy density at the Earth's orbit. The interstellar gas in the Local Interstellar Cloud (LIC) wherein our heliospheric system is embedded has an average density of $N_{\rm LIC}\approx 3\times 10^5 \,{\rm m^{-3}}$ and a relative velocity $V_{\rm LIC}\approx 26\,{\rm km/s}$ with respect to the solar system. Its kinetic energy density is thus ${\cal E}_{\rm LIC}\approx 2\times 10^{-13}\,{\rm J/m^3}$. Equating ${\cal E}_{\rm SW}(R)={\cal E}_{\rm LIC}$ one immediately finds that the upwind boundary of the region which is dominated by the solar wind, the heliopause, is to be expected around a heliocentric distance of $R_{\rm HP}\approx 130$ AU. Downwind the heliopause will be at a distance somewhat farther away with the heliosphere probably forming an extended tail. 

There are many uncertainties in this estimate but for the solar wind velocity. One of them is that the LISM does not consist just of neutral atoms; it is partially ionised by the UV radiation of nearby stars. In addition it contains a small amount of energetic Cosmic Rays \citep[a recent collection of the composition of the LISM can be found in][see also Table\,\ref{chapTS-table1}]{Izmodenov2006}. Moreover, the LISM is weakly magnetised as well. These components and the presence of the weak interstellar magnetic field affect the interaction between the solar wind and the LISM. In particular the ionic component turns out to be important in this interaction. 
\begin{table}[t]
\caption{Properties of the Local Interstellar Medium in the Local Interstellar Cloud$^*$}
\begin{center}
\begin{tabular}{lcl}
\hline
& & \\
Parameter & Observation/estimate & Method \\
& & \\
\hline
& & \\
Sun/LIC relative velocity & $26.3\pm 0.4$ & He atoms$^1$ \\
  (km\,/\,s)                                           & $25.7$ & absorption lines, Doppler effect$^2$ \\ [1ex]
Temperature & $6300\pm 340$ & He atoms$^1$ \\ 
   (K)                     & $6700$ & absorption line broadening$^2$ \\ [1ex]
 $N_H$ (atomic hydrogen, ${\rm m^{-3}}$)& $(2\pm 0.5)\times10^5$ & pick up ions$^3$ \\ 
 $N_i$ (protons, ${\rm m^{-3}}$)& $(0.3 - 1)\times10^{-5}$ & pick up ions$^3$ \\ [1ex]
 $B_{\rm IS}$ (magnetic field, nT)& $0.2-0.4$ & direction$^4$: $30^\circ<\alpha<60^\circ$ \\ [1ex]
 CR-pressure $\textsf{P}_{\rm CR}\,({\rm MeV/m^3})$& $\sim0.2$ & \\
 & & \\
 \hline
 %& & 
 \end{tabular}\end{center}
{\footnotesize $^*$after \cite{Izmodenov2006a}, ~~$\alpha=\angle({\bf B}_{\rm IS},{\bf V}_{\rm IS})$. \\ $^1$\cite{Witte2004}, \cite{Mobius2004}, $^2$\cite{Lallement1996}, $^3$\cite{Gloeckler1996}, \cite{Gloeckler1997}, \cite{Geiss2006}, $^4$\cite{Lallement1995,Lallement2005}.}
\vspace{-0.3cm}
\label{chapTS-table1}
\end{table}%

Nevertheless, from the above estimate one obtains an impression on the order of magnitude of the global size of the region that is dominated by the solar wind. One of the uncertainties is the thermal pressure of the solar wind. With increasing heliocentric distance the number density of neutral gas atoms which leak in across the heliopause increases. These neutral gas particles, predominantly atomic hydrogen, undergo multiple charge exchanges with the fast solar wind protons whereby they transform into cold pickup ions which couple to the solar wind and are heated up to four times the solar wind kinetic energy, a mechanism that has been described by \cite{Winske1985} as the result of the interaction of the freshly injected and accelerated pickup ion ring distribution with electromagnetic instabilities. They heat the solar wind and at the same time retard it. At the same time they produce fast streaming neutral hydrogen atoms flowing outward at solar wind speed. This process can repeat itself generating a substantial component of these pickup ions in the outer heliosphere which carry a substantial fraction of energy. When interacting with the Termination Shock a fraction of these pickup ions is believed to be accelerated to comparably high energies \citep{Fisk1974,Pesses1981} and forms the so-called Anomalous Cosmic Ray (ACR) particles in the solar wind. In the outer heliosphere the retardation and heating effect on the solar wind and the pressure that is attributed by the pickup ions and the Anomalous Cosmic Rays is believed to be quite substantial and modifies the properties of the outer heliospheric solar wind. 

Another uncertainty is found in the behaviour of the interplanetary field the field lines of which follow geometrically the Parker spiral. At 1 AU the angle of the parker spiral is roughly 45$^\circ$ against the radial direction, rapidly increasing with radial heliocentric distance. In the outer heliosphere it is believed to be close to 90$^\circ$. In the ecliptic plane the magnitude of the interplanetary magnetic field decreases $B\propto R^{-1}$. Thus the ratio of the kinetic to magnetic pressure $\beta_{kin}=2\mu_0{\cal E}_{\rm SW}/ B^2=$\,const is roughly constant. Since at 1 AU $\beta_{kin}>1$, the solar wind remains high $\beta_{kin}$ plasma also in the outer heliosphere (see Table \ref{chapTS-table2}) such that the magnetic field becomes unimportant for the processes at the heliopause. This holds, of course, only as long as the solar wind is not substantially retarded by the presence of a strong Anomalous Cosmic Ray component. Table \ref{chapTS-table2} also gives the thermal, pickup ion and Anomalous Cosmic Ray pressures and kinetic to magnetic pressure ratios. In pressure pickup ions are comparable to the dynamic solar wind pressure thus completely dominating the ion-plasma $\beta$ in the comoving plasma frame. It is only the pickup ion $\beta$ that has to be taken into account in all the internal plasma processes like instabilities, wave excitation and wave-particle interactions, with a contribution of the Anomalous Cosmic Ray component.
\begin{table}[t]
\caption{Densities and pressures of solar wind components as 80 AU}
\begin{center}
\begin{tabular}{lccc}
\hline
& & &\\
Component & $N\,\,({\rm m^{-3}})^*$ & $\textsf{P}\,\,({\rm MeV/m^3})^*$ & $\beta=2\mu_0 \textsf{P} /B^2$ \\
& & &\\
\hline
& & &\\
SW protons & $(7-15)\times10^2$ & $10^{-3}-10^{-4}$ (thermal) & $7\times 10^{-3}$ \\
                                             & & $\sim 0.5-1.0$ (dynamic) & $50-100$ \\ [1ex]
Pickup ions               & $\sim2\times 10^2$ & $\sim 0.15$ & $15-20$\\ [1ex]
ACR  & & $0.01-0.1$ & $ 1-10$\\ [1ex]
 \hline \\[-0.7ex]  
\multicolumn{2}{l}{{\footnotesize $^*$after \cite{Izmodenov2006a}.}}& & 
 \end{tabular}
\end{center}
\vspace{-0.5cm} 
\label{chapTS-table2}
\end{table}

In addition the interplanetary magnetic field exhibits a sector structure with oppositely directed magnetic fields. Since the spiral fields have roughly 90$^\circ$ angle when going along the radial direction outward, one alternately passes through adjacent regions of alternating field directions or, in other words, flows across a washing board formed by the undulating interplanetary current sheet. 

The magnetic field at 100 AU, on the other hand, is just a factor 100 less than at 1 AU. This value is about $B_{\rm HP}\sim 5\times10^{-2}$ nT, a value that is just below the value of the interstellar magnetic field (cf. Table\,\ref{chapTS-table1}). This situation is similar to that near Earth where the solar wind also carries a weaker magnetic field than the geomagnetic field in the magnetosphere, the obstacle with that the solar wind interacts at Earth. However, when taking into account a presumable compression factor $B_2/B_1\sim (2-4)$ at the Termination Shock, the heliospheric magnetic field near the heliopause becomes the same order as the interstellar magnetic field $B_{\rm IS}$. Since at the heliopause both fields are inclined, they can in principle reconnect and thus can lead to a  much closer coupling between the solar wind and the interstellar medium than naively believed, causing asymmetries in the geometrical form of the heliopause, leakage of plasma, generation of plasma jets and other effects.

\section{\hspace{-0.0cm}The Argument for a Heliospheric Termination Shock}
\noindent Given these conditions and uncertainties one of the most interesting effects in the outer heliosphere is  the formation of a shock wave that, on the global scale, surround the heliosphere, the heliospheric Termination Shock (or Termination Shock). It is believed that it surrounds the entire heliosphere because the interstellar gas can theoretically be considered as being at rest compared to the fast solar wind flow. 

Its formation is understandable if one considers the super-magnetosonic nature of the solar wind. 
The constancy of the dynamical $\beta_{kin}\equiv{\cal M}_A^2={\rm const}\gg 1$ implies that the solar wind in the outer heliosphere remains to be a super-Alfv\'enic and supercritical plasma flow. The  violent interaction of the supercritical solar wind flow with the interstellar gas necessarily leads to the formation of a shock inside of the heliopause where the solar wind stream is retarded to sub-magnetosonic speed and is heated. 

The location of this Termination Shock can be estimated from the formula for the stand-off distance of a bow shock in front of the blunt obstacle. Clearly the heliopause is a blunt though concave obstacle, a circumstance that yields an inward moving (in the solar wind frame) Termination Shock of as well concave shape. The distance $\Delta_{\rm TS}$ of this Termination Shock from the heliopause is given by the empirical formula that holds for bow shocks
\begin{equation}
\frac{\Delta_{\rm TS}}{R_{\rm HP}}\approx 1.1\frac{N_1}{N_2}
\end{equation}
where $N_1$ is the total solar wind density inside the heliosphere, and $N_2$ is the total plasma density outside the Termination Shock in the turbulent heliosheath. The Termination Shock is the bow shock of the local interstellar gas obstacle in the solar wind. 

Again, for a strong shock with $N_2/N_1=4$ the distance of the Termination Shock from the heliopause becomes approximately $\Delta_{\rm TS}\approx 36$ AU. Expressed as the distance from the sun this yields a heliocentric distance of $R_{\rm TS}\approx 94$ AU. On the other hand, if the shock becomes weaker -- which is the case when $N_1$ increases due to the addition of a substantial fraction of pickup ions in the outer heliosphere -- then the nominal location of the Termination Shock should shift inward into the heliosphere to a position substantially closer to the sun. Detecting the location of the heliospheric Termination Shock is therefore an indication of the strength of the shock respectively of the state of it being mediated by pickup ions and the pressure of Anomalous Cosmic Rays. The estimated solar wind parameters for the four dominant solar wind components, thermal and dynamic solar wind protons, interstellar pickup ions, and Anomalous Cosmic Ray particles are given at a distance of 80 AU in Table\,\ref{chapTS-table2}. The most intriguing observation is that the pickup density and pressure is comparable to the solar wind density and dynamic pressure, while the ACR pressure may also contribute up to 20\%. 

The region outside of the Termination Shock, i.e. the region between the Termination Shock and the heliopause, is the Heliosheath. By its nature as being the transition region between obstacle and shock it should be a highly turbulent medium of compressed density and magnetic field, of increased temperature and in the ideal case zero normal flow velocity as the simple model described here assumes that there is practically no solar wind flow across the heliopause into the interstellar gas. The solar wind should be transformed into a flow that is tangential to the heliopause  surrounding the solar system in the heliosheath boundary layer while slowly mixing into the surrounding local interstellar medium. This mixing is provided by the weak collisions and charge exchange processes between the solar wind and the interstellar atoms and, if the magnetic field becomes important, is also caused by reconnection at the heliopause. 
\begin{figure}[t!]
\centerline{\includegraphics[width=1.0\textwidth,clip=]{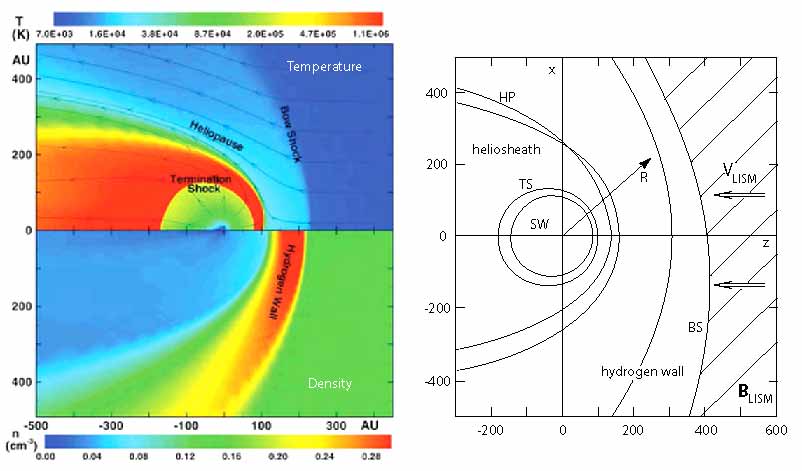}}
\caption[Izmodenov]
{\footnotesize The heliosphere. {\it Left}: Global kinetic simulations \citep[from][]{Zank2003} of the interaction of the heliosphere and the local interstellar medium. The upper half of the figure shows the kinetic temperature distribution in the heliosphere and local interstellar medium, the lower part the density distribution. The ion component of the local interstellar medium and leaking solar wind ions accumulate in the dense hydrogen ion wall. The boundary is the heliopause. A bow shock in the interstellar medium develops when the medium is braked by the heliosphere obstacle. The solar wind is compresssed into the heliosphere. Inside the heliosphere the Termination shock evolves when the solar wind is stopped. This simulation is without interstellar magnetic field and is thus symmetric. {\it Right}: The effect of the interstellar magnetic field on the heliosphere is shown as the dark lines compared to the light lines without magnetic field \citep[after][]{Izmodenov2006a}. No solar wind magnetic field is assumed. The magnetic field in the LISM deforms the heliosphere system and compresses the heliopause and Termination shock. The values used in this simulation are: $N_{i,{\rm LISM}}=6\times 10^4\, {\rm m}^{-3}, \,\,N_{\rm H,LISM}=1.8\times 10^4\,{\rm m}^{-3}, \,\,B_{\rm IS}=0.25\,\,{\rm nT}$. } \label{chapTS-fig-mueller}
\end{figure}

\section{The Global Heliospheric System}
\noindent Since the heliosphere has a radius (upwind heliopause distance) of $R_{\rm HP}\sim 100$ AU its size is huge. The collisional mean free path is only a few AU (the dominant collisional process being charge exchange with neutrals from the LISM with cross section $\sigma_{\rm ex}\sim 10^{-18}\,\,{\rm m}^2$), still large compared to terrestrial dimensions but short compared with the dimensions of the heliosphere. It is thus believed that it can be described sufficiently well within a hydrodynamic or magnetohydrodynamic approach. This is certainly true what concerns its gross dimension and structure but does not apply to the interesting local properties of the Termination Shock which -- as we know from the previous chapters --  has dimension of the ion inertial scale $\lambda_i=c/\omega_{pi}$ with substructure on the electron inertial scale $\lambda_e$ where the electron physical processes are taking place.  From Table\,\ref{chapTS-table2} we find that $\lambda_i\sim 170\,\,{\rm  km}\approx 10^{-6}$ AU. This is much less than the collisional mean free path which is of the order $\sim (50-100)$ AU, identifying the Termination Shock as a truly collisionless supercritical and very thin shock transition. Nevertheless, its gross shape can be described by fluid theory as it also holds for the planetary bow shocks and there is no good reason to not extrapolate to the Termination Shock as well. 

Figure\,\ref{chapTS-fig-mueller} taken from \cite{Zank2003} shows the results of multi-fluid simulations of the outer heliosphere in the local interstellar medium of the local interstellar cloud (so-called kinetic simulations as they include the multi-component nature of the solar wind and LISM coupling them via collisional interaction including charge exchange). In the left part of the figure the magnetic field is neglected. Thus the system is symmetric, and only a half plane is given space on top for the temperature ($T$ in K), below for the density ($N\to n,\,{\rm in\,\,cm^{-3}}$). The colour code means that red is high, blue is low. Correspondingly, the high kinetic energy in the solar wind transforms into high temperature. 

The upper part of the figure on the left very clearly shows the formation of the heliosphere in the interaction to which the solar wind is confined. The inner part of the heliosphere is bound to the greenish-yellow region lying inside the Treminal shock (called here Termination shock) which has an approximately ellipsoidal shape with the sun in one of its foci. Outside the Termination Shock in the region of the broad heliosheath the medium is hot until reaching at the heliopause boundary where the flow (blue lines) is diverted towards the flanks and where it mixes with the LISM and cools. This mixing regions widens at the flanks of the heliosphere. 

On the other hand, outside the heliosphere the LISM is also heated over a broad range of distances of the order of $\sim 100$ AU width with the flow lines of the LISM are bent around the heliosphere. Note that inside the Termination Shock the yellow colour indicates that the solar wind in its outer parts is hot which is due to the presence of the pickups. It is also worth noting -- though not of particular importance for the physics of the Termination Shock -- that in these simulations a bow shock forms in the LISM far in front of the heliopause. The hydrogen wall turns out as the downstream transition layer from the LISM to the heliosphere in this interaction. The reason for this bow shock (the existence of which is still hypothetical and the physics of which is different from magnetised planetary bow shocks being more similar to the Venusian or cometary bow shocks) is that the LISM in this model simulation is supersonic and the interaction with the heliosphere is thus shocklike.  

The lower part of the figure shows the density which is low inside the heliosphere but shows a huge enhancement in the upwind pile up region at the nose of the heliosphere where the LISM ions are compressed and accumulated. this is called here the Hydrogen Wall. Leakage of the LISM into the heliosphere is indicated by the erosions on the outer heliosphere.

The right part of the figure, redrawn after \cite{Izmodenov2006a}, shows the effect of the presence of an interstellar magnetic field on the shape of the heliosphere and Termination Shock. The light lines correspond to the simulations on the left without magnetic field, the heavy lines include the LISM magnetic field effect. The main effect in this simulation -- which does not include a solar wind magnetic field and thus also no reconnection effects -- is the deformation by the additional pressure of the LISM magnetic field this deformation, on the scale shown, is very small on the TS but strong on the heliopause and heliospheric bow shock wave.
\begin{table}[t]
\caption{$\quad$Ion Reflection Ratios $R=N_{ref}/N_{in}$ at Quasi-Parallel Shocks$^*$ }
\begin{center}
\begin{tabular}{l|cc|cc|cc}
\hline
& & & &&&\\[-0.3ex]
&\multicolumn{4}{c|}{Stationary shock}&\multicolumn{2}{c}{Travelling shock}\\[0.3ex]
${\cal M}_A$ &\multicolumn{2}{c|}{5.3}  &\multicolumn{2}{c|}{10.4} &\multicolumn{2}{c}{10.0}\\[1.5ex]
\hline
& & & & \\
Ion & Pickup& Diffuse& Pickup& Diffuse& Pickup& Diffuse\\[0.3ex]
& \multicolumn{6}{c}{(\%)}\\ [1ex]
\hline
& & & &&& \\[-0.3ex]
H$^+$ & 42 & 0.9 & $\sim$\,29& &10.3&\\
&&&&&&{$<0.6$}\\ 
 He$^+$ &25 & 0.6 & $\sim$\,31& &6.2&\\ [1ex]
 \hline 
\multicolumn{7}{l}{}\\[-0.7ex]
\multicolumn{7}{l}{{\footnotesize $^*$($\thetabn=5^\circ, \beta_i=\beta_e=0.5$, $N_{H^+}/N=10^{-3}$, 5\% He$^{2+}$)}}\\[-0.3ex]
\multicolumn{7}{l}{{\footnotesize ~~data taken from \cite{Scholer1999a}}}
 \end{tabular}
\end{center}
\vspace{-0.5cm} 
\label{chapTS-table3}
\end{table}
\section{Termination Shock Properties:  Predictions}
\noindent Calculation like those in the above simulations can provide only gross properties of the Termination Shock, its presumable location in the heliosphere and its presumable shape. Even this information is questionable as long as no solar wind magnetic field has been included. The latter is subsequently imposed, behaving passively, non-dynamically, and having the Parkerian (Archimedean) spiral form which implies that it is about tangential to the nominal Termination Shock shape such as it follows from the simulations. Hence the Termination shock is, in terms of these simulations, a supercritical, almost perpendicular, collisionless shock. It might, however, be modified to some extent by the presence of the energetic and comparably dense pickup ion component. Thus the Termination Shock should be a huge shock surface of a quite strong collisionless shock, by far the most extended shock wave in the entire heliosphere.

All these claims are rather uncertain as long as they cannot either be supported (or falsified) by direct observations (as will be discussed in the next Section) or made plausible with the help of suitable numerical simulations on scales which are appropriate for the study of the Termination Shock microphysics. In particular, the claim that the Termination Shock is strictly perpendicular might be mistaken as it is based on information that is extracted from global scales of the order of $\sim0.3$ AU only, the presumable average distance between the solar wind corotating magnetic sector structures. Even at the Earth's orbit at 1 AU the direction of the solar wind is quite unstable and is only in the long term and on the large scale average aligned along the Parker spiral. This variability is usually attributed to the proximity of the Sun and the variability of the inner heliospheric solar wind. There is, however, little reason to be certain about the quietness of the outer heliospheric solar wind and the greater stability of its local magnetic field. In fact, Pioneer 11 observations at 35 AU \cite{Smith1993} suggest a variation in the magnetic field direction around the Parker spiral of $\pm25^\circ$.

The presence of the dense pickup ion shell distributions in the outer heliosphere should cause a high level of Alfv\'enic turbulence of long wavelengths electromagnetic ion-cyclotron waves which in the collisionless cold outer solar wind plasma with its long scales are only very weakly damped and are not hindered to steepen until reaching large amplitudes. Their transverse character implies that the assumption on the outer heliospheric solar wind magnetic field of being perpendicular to the direction of the solar wind flow ceases to hold. At the contrary, on the solar wind frame of reference the wave magnetic field is transverse to the magnetic field and might thus have a strong component in the direction perpendicular to the spiral. In this case, the Termination Shock will -- on the local scale -- behave much more like a quasi-parallel shock than a quasi-perpendicular or even strictly perpendicular shock. In addition, even when the field is of spiral form is there no reason to believe that the Termination Shock is tangential to the field. The field lines might cross the shock at its one flank and re-enter it on the other flank if the shapes of the magnetic field spiral and Termination Shock do not align which will be the case when the magnetic field behaves completely passively and the interaction between the solar wind and the LISM is due to the flow as is assumed in the above global simulations. 

It has  been concluded by \cite{Liewer1993} from hybrid simulations including 10\% pickup protons at quasi-perpendicular shocks that below a shock normal of  $\thetabn=60^\circ$ pickup protons are very efficiently backscattered from the shock and generate upstream waves which cause pitch angle diffusion on the particles. The scattering ceases at larger angles, however. Although the Termination Shock is expected to be dominantly perpendicular, \cite{Kucharek1995} investigated pickup ion reflection for the case that the Termination Shock would temporarily have quasi-parallel character. In contrast to the \cite{Liewer1993} simulations they included heavier ions suspecting that they are backscattered even stronger than protons. Their simulation settings are one-dimensional hybrid including mass ratio weighted H$^+$ and He$^+$ ions and  5\% He$^{2+}$, Mach number ${\cal M}_A\approx 8$, investigating the two cases of $\thetabn =50^\circ$ and $70^\circ$ thereby confirming the above \cite{Liewer1993} result of backscattering only for angles $\lesssim 60^\circ$. The initial pickup distributions are accelerated spherical shells. The heavier ions have been treated as test particles in a non-selfconsistent way. The result of this simulation is that with increasing mass the reflection of pickups decreases. This can be simply explained by the larger energy of the particles which can overcome the shock ramp potential. Applied to the Termination Shock this implies that for oblique shock normal angles the Termination Shock can well act as a good reflector of pickup ions thus providing a seed population for their further acceleration when they return to the shock. 
\begin{figure}[t!]
\centerline{\includegraphics[width=1.0\textwidth,clip=]{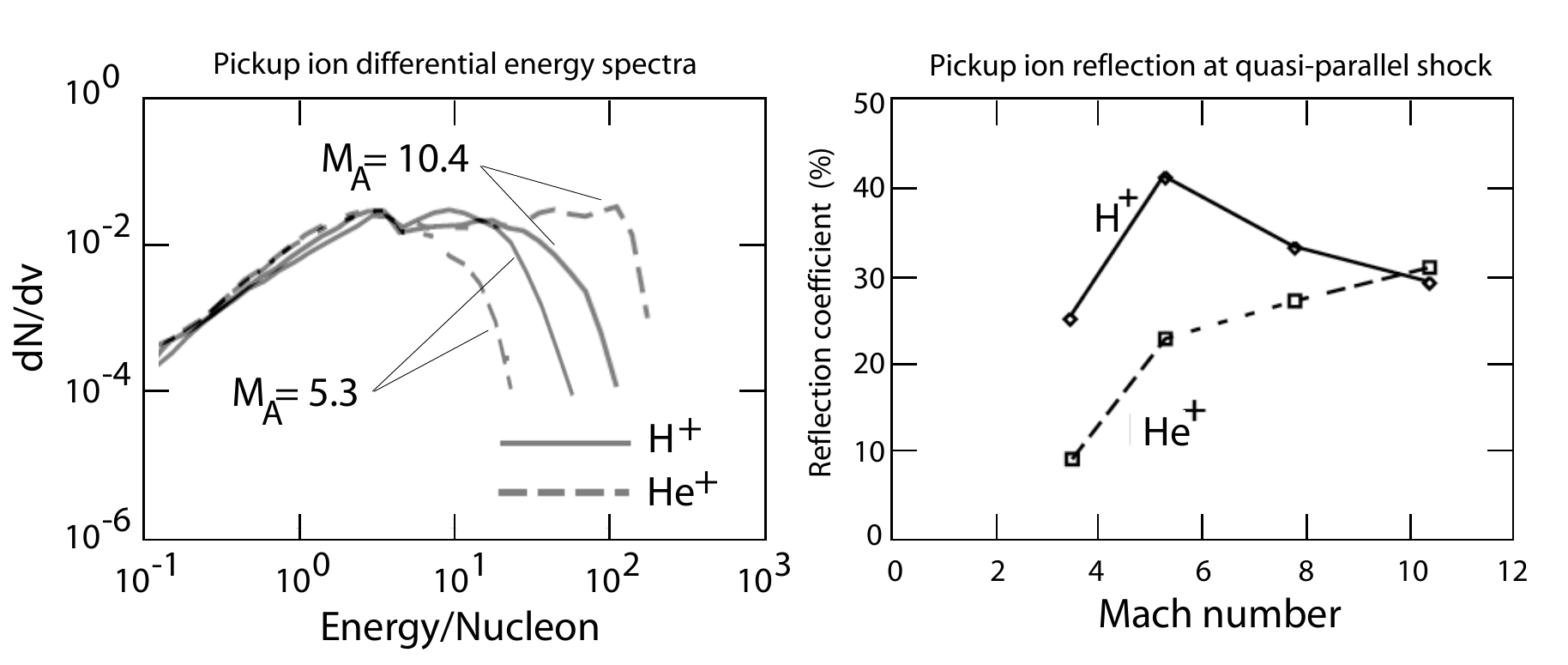}}
\caption[Scholer-Kucharek]
{\footnotesize Hybrid simulations of pick up ion reflection at quasi-parallel shocks ($\thetabn=5^\circ$) in view of acceleration of Anomalous Cosmic Rays at the Termination Shock \citep[data from][]{Scholer1999a}. {\it Left}: The differential energy spectra of the pickup ions for two different Mach numbers of ${\cal M}_A=5.3$ and 10.4 (in units of the upstream kinetic energy) . The low energy spectra are identical but the higher Mach numbers yield higher  heavy pickup ion energies. At the low Mach number proton pickups reach higher energy. At the high Mach number the He$^+$ ions are accelerated stronger than protons. {\it Right}: The reflection coefficients as function of Mach number. Protons reach maximum reflexivity around medium Mach number while the reflection coefficient for He$^+$ pickups still increases and overturns the proton coefficient at about Mach number 10. Pickup reflection coefficients are in the range of tens of percent, two orders higher than the shock reflexivity of diffuse ions.}\label{chapTS-fig-schku}
\end{figure}

This has been further investigated by \cite{Scholer1999a}  who right away assumed a quasi-parallel shock in their one-dimensional hybrid simulations of  $\thetabn=5^\circ,  {\cal M}_A=5.3$ and ${\cal M}_A=10.4$ shocks in view of the generation of the Anomalous Cosmic Rays (see Table \ref{chapTS-table3}). They adopt the picture that pickup ions are in a first step predominantly accelerated by travelling interplanetary shocks in the region farther inside the heliosphere, and when transported out to the Termination Shock become accelerated there to the high Anomalous Cosmic Ray energies by efficient reflection at the stationary Termination Shock and the action of assumed non-linear trapping and non-adiabatic acceleration. Again the heavier ions than pickup protons are treated as test particles. The main results of these simulations are shown in Figure \ref{chapTS-fig-schku}. The particle distributions (differential energy fluxes) are shown on the left. The particle reflection coefficients are shown on the right as function of the Mach number. The authors find reflection ratios $R=N_{ref}/N_{in}$ for pickup protons of 42\%, for He$^+$ of 25\% compared to the respective shock generated 0.9\% and 0.6\% diffuse proton and He$^+$ ions in the foreshock. High Mach numbers imply higher final energies of the heavier ions (see the left part of he figure). This can clearly be explained only when the non-adiabatic behaviour of the pickup ions at the shock dominates over the shock potential effect at the higher Mach numbers. This can be understood  because the amplitude of the shock produced waves increases with Mach number, as had been shown by \cite{Scholer1997}.  

It follows that shocks can indeed inject substantial amounts of pickup ions into the Fermi cycle, for the pickups this fraction is much higher than the fraction of shock produced upstream particles. This is due to the very low velocities of the pickups in the shock normal direction since they have merely been accelerated into the direction of the tangential upstream motional electric field perpendicular to the shock normal. Thus their energy is low enough for being reflected from the shock potential and trapped in the field close to the shock for a while when becoming non-adiabatically accelerated to high energies, a process that is known from shock surfing. The distribution of these reflected and accelerated pickup ions has the shape of a pancake with more energy in the perpendicular direction. Such a component excites two types of waves: anisotropy driven whistlers and ion-ion instabilities when back entering the upstream solar wind. 

Applied to the Termination Shock which is located in the midst of the highly increased fraction of interstellar pickup ions this implies that the Termination Shock could be an excellent place to reflect and further accelerate those pickup ions -- protons and heavier less abundant elements -- to become Anomalous Cosmic Rays. The condition for this is that the Termination Shock is not an ideally quasi-perpendicular shock but has regions where locally the shock normal angle to the interplanetary magnetic field is substantially smaller than $\thetabn\lesssim 60^\circ$. Such regions are expected to exist possibly at its flanks where in the idealised picture the spiral interplanetary heliospheric magnetic field should cut through the Termination shock at some angle. However, they can also be generated by surface instabilities of quasi-perpendicular shocks which are driven by the gyrating reflected ions in their accelerated motion in the upstream motional electric field along the shock surface \citep{Lembege1992,Lowe2003,Burgess2007}. Such an instability resembles a flow instability of the Kelvin-Helmholtz family but is entirely due to the kinetic effect of the reflected ions. It requires a supercritical shock and sufficiently low $\beta_i$. When it sets on, the surface of the quasi-perpendicular shock ceases to be smooth but develops ``ripples" which locally affect the shock normal direction. 

More recent full particle PIC simulations performed for quasi-perpendicular shocks by \cite{Scholer2003} with real mass ratios $m_i/m_e=1836$ have investigated the $\beta_i$ dependence of shock reformation, ion reflection, and the cross-shock potential and have found that the shock potential divides between the foot and the ramp and is restricted to scales of $\sim 4 \lambda_e$. At the low thermal ion-$\beta_i$ that is expected at the quasi-perpendicular Termination Shock coherent ion trapping and acceleration of pickups as required in shock surfing mechanisms is thus improbable. If this turns out to be valid then the acceleration of pickups at the Termination Shock will mainly be due to acceleration in those regions where the character of the Termination Shock is quasi-parallel and should thus lead to anisotropies in the spatial distribution of Anomalous Cosmic Rays and pancake forms in the energetic ion velocity distributions. There are still a number of caveats in this conclusion which is based on the one hand on hybrid simulations, on the other hand on just one-dimensional PIC simulations still leaving much space for interpretations, speculations and also surprises.

On the other hand the same simulations have also shown that quasi-perpendicular shocks like the expected Termination Shock cannot be stationary but undergo reformation at low $\beta_i<0.4$ at the simulation Mach number ${\cal M}=4.5$ unless the pickup ions in the outer heliosphere contribute to such an increase in $\beta_i$ that reformation ceases to occur.  This effect might be partially compensated by a low plasma-to-cyclotron frequency ratio $\omega_{pe}/\omega_{ce}\sim 1$. 
From Table \ref{chapTS-table2} we have that in the foot region of the Termination Shock $\omega_{pe}/\omega_{ce}\sim 1.6$,  even if all plasma components including pickup ions are taken into account (remembering that the ratio $N/B^2$ is independent of radius in the expanding solar wind). This favours reformation even though the high pickup $\beta$ in Table \ref{chapTS-table2} tends to suppress it when the pickup ions are involved in the shock formation. The large mass ratio simulations also show that real reformation of quasi-perpendicular shocks does not so much depend on the number of gyrating ions than on the electromagnetic modified two-stream instability in the foot of the shock. In this case the $\beta-i$ dependence is not sufficiently well known and the shock might reform anyway. 
\begin{figure}[t!]
\centerline{\includegraphics[width=0.85\textwidth,clip=]{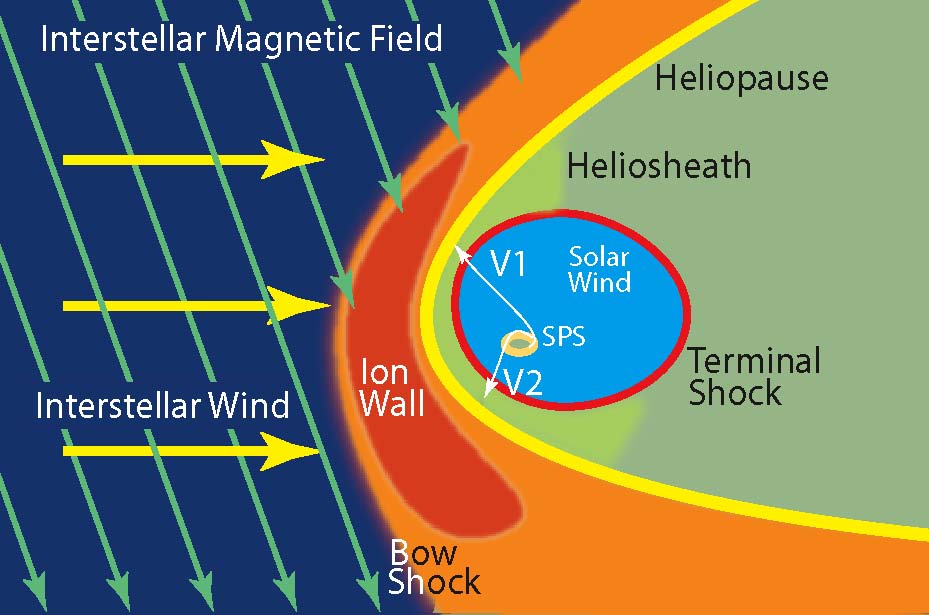}}
\caption[TS]
{\footnotesize Sketch of the upwind heliosphere in the magnetised interstellar wind blowing against the solar wind, creating a Bow Shock and Ion Wall in the LISM, and the Heliopause. SPS is the solar planetary system showing the two orbits of the Voyager 1 (V1) and Voyager 2 (V2) spacecraft. V1 is shown to cross the Termination Shock (TS) at larger northern distance than V2 in the south. The Termination Shock is the red ring that confines the solar wind. The (exaggerated) asymmetry of the TS and location of SPS in the solar wind is caused by the interstellar magnetic field. Outside of the Termination Shock and the heliopause is the Heliosheath. Colour symbolises density, dark blue the low LISM density up to blue the low solar wind density. Density gradients are not included. }\label{chapTS-fig-TS}
\end{figure}

It is thus probable that the Termination Shock is rather a non-stationary shock than a stationary quasi-perpendicular shock. Locally it will be oscillating back and forth around its nominal position, a process that is driven by the variations in the interplanetary solar wind medium and its magnetic field direction and sector structure. It necessarily leads to surface waves running along the shock. The Termination Shock can thus be expected to exhibit quasi-periodic or also irregular variations in its $\thetabn$ and the direction of its normal with respect to the magnetic field. Because of the high pickup ion density near the Termination Shock one might even speculate that the capability of the Termination Shock of reflecting  a fraction of hot pickup ions might cause shock reformation on the scale of a pickup ion gyro radius that is produced by accelerated gyrating reflected pickup ions and involves an instability between the pickup ions and the upstream electrons of the main flow.

\section{Observations: The Voyager Passages}
\noindent There is no way to observe the Termination Shock from Earth. With one exception, even from near Earth space the Termination Shock cannot be observed in any radiation. The exception is the Anomalous Cosmic Ray component. It is called anomalous because it has little to do with the real Cosmic Rays from the Galaxy. Though it is still not approved, it is believed that these energetic particles are coming from the Termination Shock where they are accelerated, being the energetic tail of the outer heliospheric pickup ions. They are thus generated inside the heliosphere having their source in the outer heliosphere. The process by which they are produced maps the composition of the LISM, not only containing hydrogen ions but also other elements. The highest abundance has been found for helium, oxygen, and neon. Anomalous Cosmic Rays therefore tell more about the LISM than the Termination Shock. Not much information can be extracted about the Termination Shock from observing them from remote because they are distributed about homogeneously in angle throughout the heliosphere. Below in the section on energetic particles we discuss the {\it in situ} evidence of their generation by the Termination Shock that is based on the Voyager passage.  

 Both Voyager 1 and Voyager 2 spacecraft on their ways out into interstellar space (see Figure \ref{chapTS-fig-TS}) ultimately crossed the Termination Shock. Voyager 1 passed it in December 2004 at radial heliocentric distance $\sim$94 AU, Voyager 2 crossed it end of August 2007 at $\sim$84 AU. The asymmetry between North and South is attributed to the action of the interstellar magnetic field and to the variations in the solar wind with solar cycle. The asymmetry of the heliosphere had been predicted earlier already from Lyman $\alpha$ observations \citep{Lallement2005} and attributed to the interstellar magnetic field direction.

Before coming to the Voyager measurements we briefly elaborate on radio observations from remote and in situ Langmuir wave measurements.

\subsection{Radio observations}
\noindent Radiation emitted from the Termination Shock has been expected to be in plasma waves and in radio waves. Intense radio emission from the outer heliosphere has indeed been observed by the Voyager spacecraft \citep{Kurth1984}. But it was soon recognised \citep{Gurnett1993} that this radiation could not have been generated at the Termination Shock unless the latter would have had very unusual properties that are not expected of any shock. Depending on the heliocentric distance of the Termination shock between 80 AU and 100 AU the local plasma density is between $5\times 10^2<N<8\times 10^2\,\,{\rm m^{-3}}$. From Table \ref{chapTS-table2} the total plasma density is of the order of $N({\rm 80\,AU})\lesssim 10^3\,\,{\rm m^{-3}}$. The local plasma frequency at the Termination Shock cannot be higher than $\omega_{pe}=2\pi\times 280$ Hz, roughly an order of magnitude below the observed radiation frequency which was in the range $1.4< f < 3.6$ kHz. In order to be emitted in the second harmonic at $ f\sim 2\omega_{pe}/2\pi$ from the Termination Shock the density at the Termination Shock should be 36 times higher than expected. The strongest shocks permit only a factor 4 density compression. Hence, the observed radio emission cannot come from the Termination Shock. Therefore \cite{Gurnett1993}  rightly attributed the radiation to come from the outer heliospheric boundary instead. The mechanism of its generation is still unclear, but \cite{Gurnett1993} and \cite{Gurnett2003} could convincingly show that its emission was related to the interaction of occasional fast solar wind streams and travelling interplanetary shocks with the heliopause \citep[see also the extended discussion in][]{Treumann2006}.

That the observed radiation has not been produced at the Termination Shock is less a surprise than the now established fact that no radiation at all has been observed that is coming from the Termination Shock. The radio emission detected by the Voyager spacecraft from the outer heliosphere on its way out were at frequency less than the local plasma frequency in the inner heliosphere being completely screened. Radiation at 300-600 Hz could not have been observed until Voyager would have passed a distance of at least 35 AU. However when Voyager was beyond this distance no emission was detected. 

The Termination Shock remained silent in the radio band even until the Voyager spacecraft had passed it after 2004. Any strong shock that is supercritical and is capable of reflecting electrons should emit radio waves, however. This holds for the Earth's bow shock, other planetary bow shocks, for solar and interplanetary Type II bursts which emanate from the corresponding coronal and interplanetary shocks; it was true even for the {\footnotesize AMPTE IRM} artificial plasma cloud injections into the solar wind \citep{Gurnett1985}. It is thus very strange that the Termination Shock which has been believed to be the strongest shock in the heliosphere is actually a radio-silent shock. Thus finding the reasons for this quiescence should illuminate the plasma conditions that prevail at the Termination Shock. It  should also contribute to the understanding of the conditions under which a shock is capable of producing radio waves. 

Current radiation theories require the presence of electron beams that propagate along the ambient magnetic field and excite Langmuir waves at the local plasma frequency. Subsequent wave-wave interactions between these Langmuir waves are believed to cause emission of radio waves at harmonics of $\omega_{pe}$. These mechanisms require either counterstreaming Langmuir waves in order to be able to satisfy the wave momentum conservation condition, or they require scattering of Langmuir waves off either thermal ions or ion acoustic waves. Some less explored theories prefer instead electron acoustic waves. But the basic condition it the presence of electron beams. These are naturally expected to be injected upstream from a sufficiently strong supercritical shock. 

Below we will see that the Voyager observations indeed have identified both the presence of electron beams as also the Langmuir waves that are excited by these beams which makes the absence of radiation even more mysterious. This absence implies that the simple merging process of either two counter streaming Langmuir or one Langmuir and one ion acoustic wave is set out of action. The first process requires generation of backscattered Langmuir waves. If the required ion acoustic fluctuations are lacking no backscattered Langmuir waves will be generated. This happens when the electron temperature is too low for ion acoustic waves to be excited by the solar wind heat flux instability, i.e. when ion Landau damping is strong. Scattering off thermal ions has weak efficiency and in the cold solar wind plasma might be suppressed. However, it remains to be unclear whether the hot and sufficiently dense pickup ions do not contribute. Possibly their densities are insufficient. Electron acoustic waves will also not be excited upstream because the electron temperatures are too low. Therefore the absence of intense radio emission from a Termination Foreshock becomes plausible. 
\begin{figure}[t!]
\centerline{\includegraphics[width=1.0\textwidth,clip=]{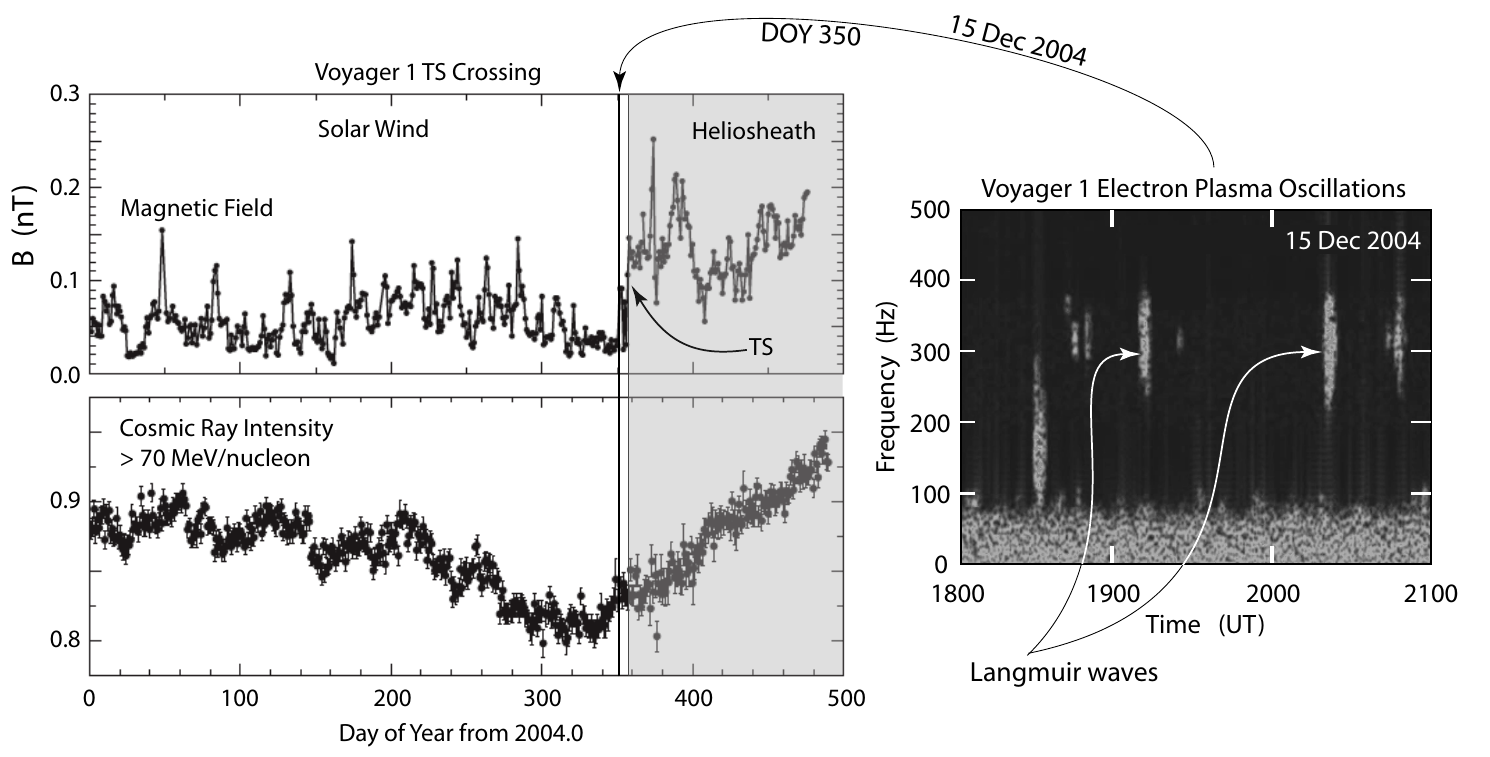}}
\caption[V1]
{\footnotesize {\it Right}: Electron plasma wave dynamic spectrogram \citep[data taken from][]{Gurnett2005} on 15 December 2004 shortly before the Voyager 1 crossing of the Termination Shock. The signals occur sporadically indicating magnetic connection of the spacecraft to the Termination Shock and the arrival of electron beams of a few keV energy. The local plasma density determined from the observation is of the order of $N\sim 10^3\,{\rm m^{-3}}$. {\it Left}: Heliospheric magnetic field magnitude showing the crossing of the Termination shock (top) and Cosmic Ray Intensity showing the unexpected continuous increase with entry into the heliosheath \citep[data taken from][]{Burlaga2005}. The vertical black line indicates the time observation of Langmuir wave spectra on the right. }\label{chapTS-fig-V1gur}
\end{figure}

\subsection{Plasma waves and electron beams}
\noindent The real question is: Why is the Termination Shock radio silent? This question is even more stringent as the Langmuir waves required for generating the expected shock radiation have indeed been observed, and that long before the Voyager spacecraft have crossed the Termination Shock, already from 11 February 2004 on during six time intervals, the last of them on 15 December 2004 at radial heliocentric distance 94.1 AU, shortly before the Voyager 1 verified and spectacular  crossing of the Termination Shock in December 2004 \citep{Gurnett2005}. 

Figure \ref{chapTS-fig-V1gur} on its right shows Gurnett and Kurth's dynamical spectrum of the electrostatic plasma waves detected with Voyager 1 on 15 December 2004 between 1800 UT and 21 UT in the frequency range from 178-300 Hz, in nice agreement with the estimated local plasma frequency upstream of the Termination Shock. Langmuir wave amplitudes of the order of $(2-3)\times 10^{-6}$ V/m have been measured, one order of magnitude above background noise. The top of the left part of the figure shows the Voyager 1 magnetic field trace with the Termination Shock crossing happening several hours after the observation of the electron plasma waves. On the day of 15 December 2004 an intense spike in the electron flux was also detected. 

It is intriguing to see the sporadicity of the electron plasma waves. Since their intensities are sufficiently far above thermal noise they can have been excited only by electron beams that flew along the outer heliospheric magnetic field and were connected to the Termination Shock. Electron beams flowing sunward along the spiral interplanetary magnetic field having energies $>26$ keV have been reported by \cite{Decker2005} to have been observed by Voyager 1 throughout the year 2004. Since this has been so for almost the full year Voyager 1 was occasionally magnetically connected to the Termination Shock from a distance of 91.0 AU over a range of $\sim3$ AU. On the day of 15 December 2004 a particularly intense spike in the electron flux was also detected.  (There are indications that is was in the foreshock of the Termination Shock already from 85 AU on which implies occasional magnetic connection over 9 AU.)  It is not known to which part of the Termination Shock the magnetic field lines were attached, initially probably more to the flanks of the Termination shock as was suggested by \cite{Gurnett2005} referring to the spiral interplanetary magnetic field. 
\begin{figure}[t!]
\centerline{\includegraphics[width=0.8\textwidth,clip=]{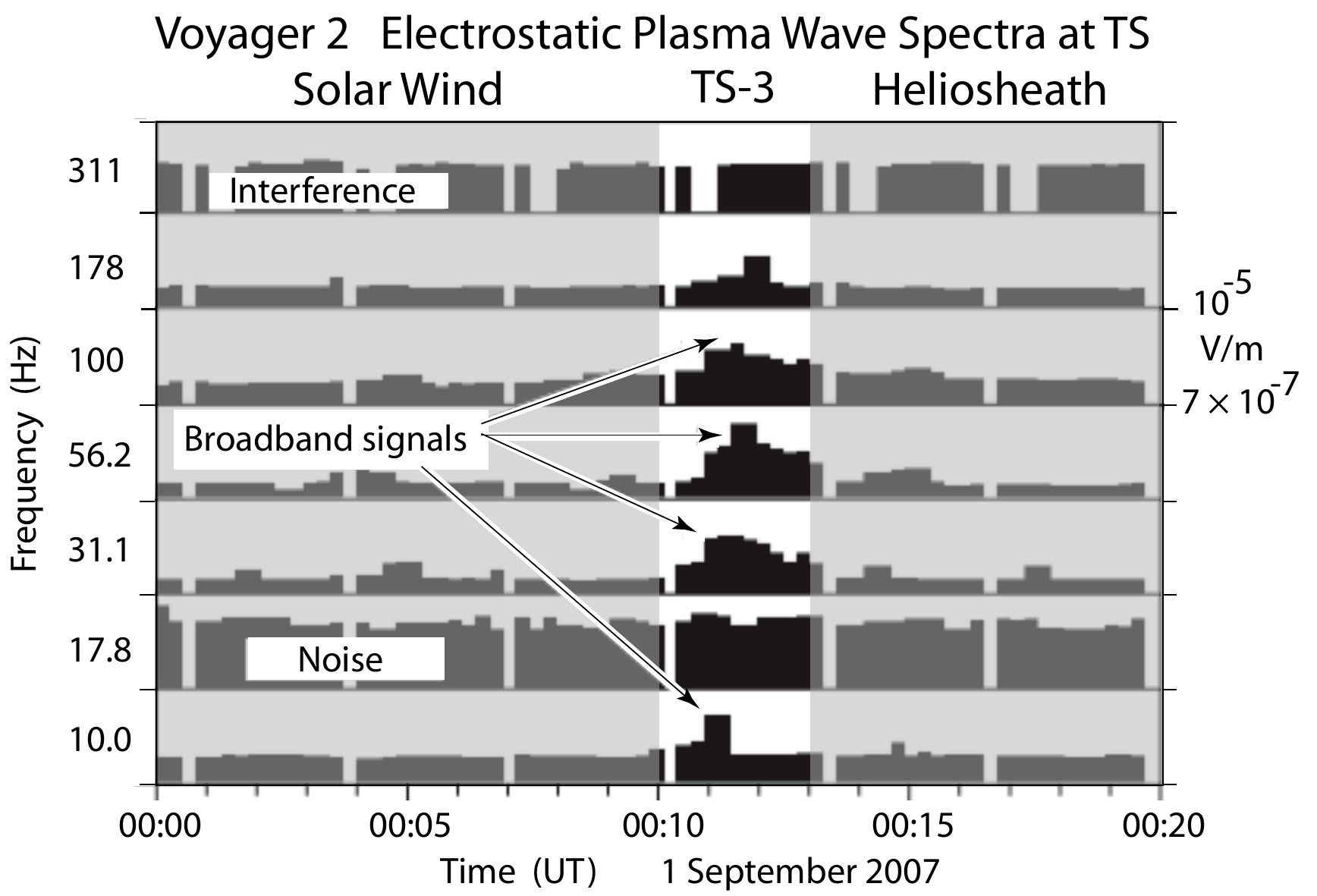}}
\caption[V1]
{\footnotesize Broadband noise signals observed during the Voyager 2 crossing TS-3 of the Termination shock on 1 September 2007 from the solar wind to the heliosheath \citep[after][]{Gurnett2008}. This signal is typical for a shock ramp crossing rsembling all bow shock crossings or interplanetary travelling shocks. It is an indication of the presence of small scale electrostatic structures in the shock transition region. The scale on the right holds for each of the 7 frequency bands. The height of the black region gives the rms electric wave field amplitude as function of time in units V/m.}\label{chapTS-fig-V2PW}
\end{figure}

The observation of the intermittence of the plasma waves also suggests that the Termination Shock cannot be strictly perpendicular, it must exhibit oscillations in the direction of its shock normal with periods of the order of hours, which \cite{Gurnett2005} suggested to be surface waves propagating along the Termination Shock. Moreover the Termination Shock must be strong enough to accelerate electron beams in the direction upstream of the Termination Shock. Since the electrons are known to be very cold at the heliocentric distance of the Termination Shock the generation of these beams could hardly have been by the Sonnerup-Wu mechanism for the electron distribution would be completely inside the loss cone. Thus if there is no mechanism that generates a hot electron halo on the outer heliospheric electron distribution function one expects that the mechanism of electron acceleration can only be due to intrinsic properties of the Termination Shock like those required by the Hoshino electron shock surfing mechanism. This implies strongly nonlinear processes to take place in the shock like the excitation of BGK modes and solitary waves. 

In this respect the most recent passage of Voyager 2 in the southern heliosphere through the Termination shock is of particular importance. Voyager 2 also observed electron plasma oscillation well before contacting the Termination Shock itself, i.e. 30 days ahead of the first crossing \citep{Gurnett2008} thus confirming the former Voyager 1 observations. From these observations an upstream (electron) plasma density of $N_e\approx 1.2\times10^3\,{\rm m^{-3}}$ was inferred. As before the presence of the Langmuir waves is convincing evidence for the capability of the Termination Shock to eject low density electron beams of keV energy along the heliospheric magnetic field upstream. Those beams were undetectable for the Voyager plasma instrumentation qualifying these Langmuir wave observations as of high diagnostic value. 

Much more important than the observation of Langmuir waves is the detection of plasma waves during crossing of the shock ramp. Examples of of the dynamic spectrum of plasma waves in such crossings have been given in Figures \ref{chap5-fig-amptesh} and \ref{chap5-fig-clusterwhisper} for quasi-perpendicular shocks and in Figure \ref{chap5-fig-ampteqpash} for a quasi-parallel shock. In every case the shock appears in the electric wave field as a broadband signal extending from the lowest frequencies up to frequencies beyond the local electron plasma frequency. The intensity of these signals is highest at the lowest frequencies. The difference between quasi-parallel and quasi-perpendicular shocks is mainly in the more turbulent behaviour of the quasi-parallel shocks where the broadband noise signals are distributed over a larger interval in space. Such signals are interpreted as the frequency space signature of a number of spatially localised electrostatic structures which cross over the spacecraft leaving their trace in a broad Fourier spectrum that is independent on the cur off frequencies which otherwise govern the dispersion curve traces of plasma waves. Because of the probably large number of solitary structures or BGK modes present in the case of a shock passage, these broadband traces are much more intense than when a single BGK mode is crossed.

\cite{Gurnett2008} have reported the observation of just such broadband electric signals in coincidence with the crossing of the Termination Shock. In fact, the Termination Shock could have been identified from these traces alone as in the outer heliosphere there would be no reason for the generation of broadband signals of this kind before entering the heliopause region. Nevertheless, the occurrence of the intense broadband noise signals coincided with plasma and magnetic field observations of the shock crossing. Of the five Termination Shock crossings identified in the magnetic and plasma data, two were also seen in the plasma wave observations, two fell into plasma wave data gaps, and one did not show a typical shock signature \citep{Gurnett2008}.  

The event preferred by \cite{Gurnett2008} is the third crossing TS-3 (their Event B).  Figure \ref{chapTS-fig-V2PW} shows the broadband signal detected during this crossing which is typical for a supercritical shock transition as described above. The scale of the rms electric wave field amplitude is given on the right. It holds for all 7 wave bands of the instrument. Note that the 17.8 Hz channel suffers from enhanced instrumental noise, and the 311 Hz channel is polluted by a gyro interference. \cite{Gurnett2008} scaled the Termination Shock spectra to similar spectra observed at the Neptune bow shock crossing finding very close similarity between both objects.

The shock crossing occurs in the unshaded time interval lasting for roughly three minutes. Prior to this  the spacecraft was in the solar wind, afterwards it is in the heliosheath. Broadband waves with amplitudes of several $10^{-6}$ V/m were detected during the shock crossing. The highest amplitudes are in the 56.2 channel well below the plasma frequency and well above the electron cyclotron frequency (which was around $f_{ce}=7.3$ Hz, embedded in the 10.0 Hz channel). From the magnetic traces and plasma observation which we discuss below it is found that the entire shock transition occurs on a scale of the order of $\Delta_{\rm TS}\sim 1 \lambda_i$, which is rather thin. On this short scale it is expected that violent nonlinear processes will take place mainly involving magnetised electrons and unmagnetised ions.  It is thus clear that the observed broadband emissions are caused by nonlinear interactions in the shock ramp.

Since the electron temperature is very low and the electron velocity in the 100 eV to keV range is well above the electron thermal speed, the responsible instability is expected to be the Buneman two stream mode which leads to BGK structures. This resembles very closely the well known behaviour of planetary bow shocks. The nonlinear wave structures are most important for the internal dynamics of the shock as well as for the trapping and acceleration of particles. Thus their identification is of crucial importance for the understanding of the behaviour of the Termination Shock which as such serves as the paradigm of a non-relativistic high Mach number supercritical shock generated in the interaction between a magnetised stellar wind and the magnetised interstellar environment. 
\begin{figure}[t!]
\centerline{\includegraphics[width=0.8\textwidth,clip=]{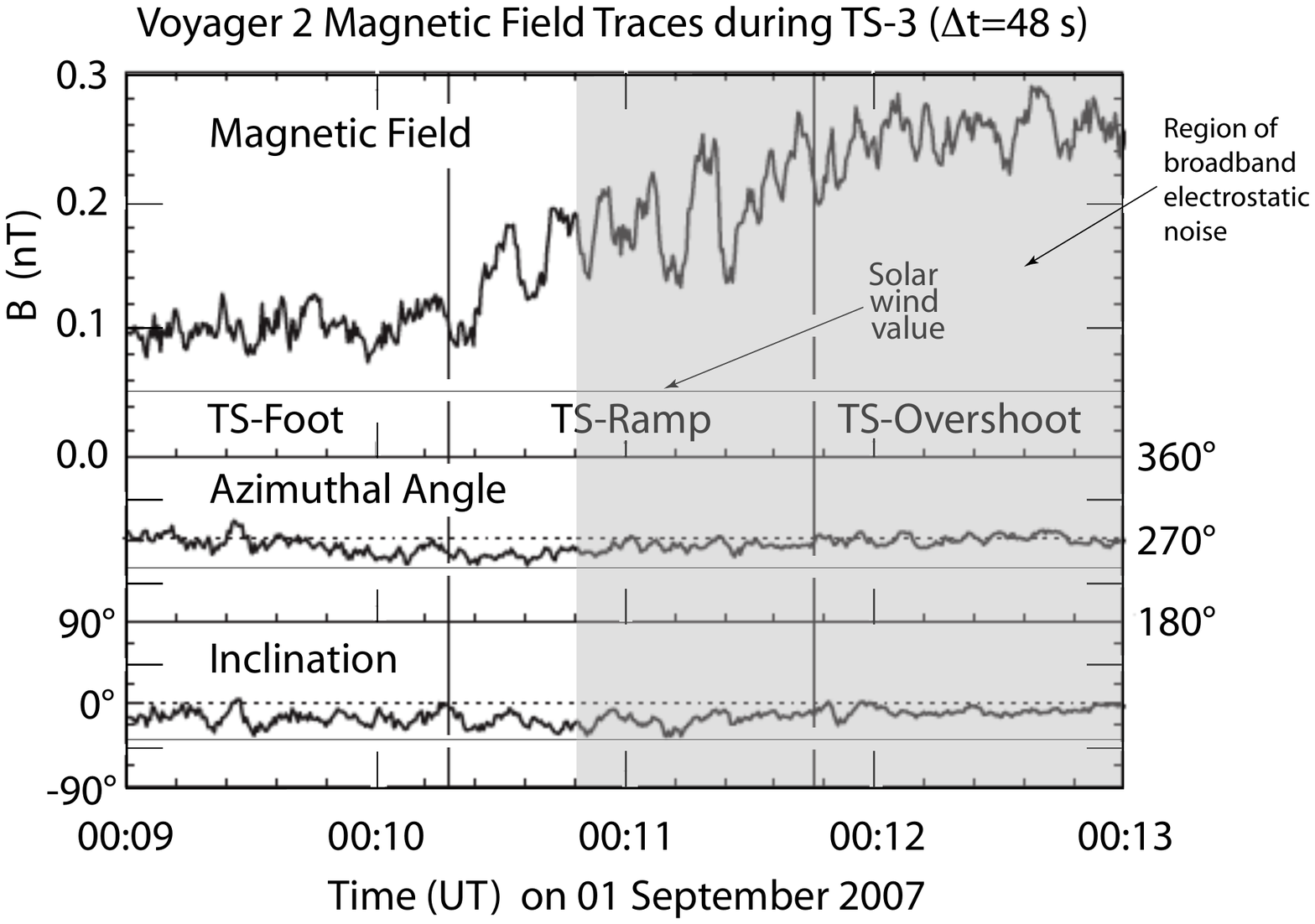}}
\caption[V2]
{\footnotesize The Voyager 2 Termination Shock crossing TS-3 in the magnetic trace \citep[after][]{Burlaga2008}. Shown is the foot region where the magnetic field is about twice the undisturbed solar wind field, the ramp regions with its strong coherent oscillations. These are from simulations know to be caused by ion trapping with the maxima in the magnetic field coinciding with the vertexes of the ion phase space rings. The transition to the overshoot is seen as incoherent fluctuations in the magnetic field. The shaded region is the time interval when the plasma wave instrument detected the broadband electrostatic noise \citep{Gurnett2008} which becomes strongest after 00:11 UT in the transition from ramp to overshoot. }\label{chapTS-fig-BPL}
\end{figure}

\subsection{Traces of plasma and magnetic field}
\noindent In situ information on plasma and magnetic field properties from the Termination Shock has been obtained recently when the Voyager spacecraft crossed it from the inner heliosphere into the heliosheath. Voyager 1 which travels at a higher speed than Voyager 2 crossed it in December 2004 at heliocentric distance of 94 AU and heliographic latitude 34$^\circ$ N after having been in the Termination Shock foreshock from 85 AU on. The slower Voyager 2 crossed the Termination Shock between 30 August 30 and 1 September 2007 at heliocentric radius 84 AU and heliolatitude 25$^\circ$ S after having been in the foreshock from 75 AU on, surprisingly finding the Termination shock substantially farther in than expected from the Voyager 1 crossing.  The distance between the spacecraft was 110 AU and 45$^\circ$ in longitude. On Voyager 1 the plasma instrument failed. Hence only magnetic field information was available to identify the shock crossing. This is shown in the Voyager 1 magnetic field trace on the left in Figure \ref{chapTS-fig-V1gur}. During the unshaded period the Voyager 1 spacecraft was in the solar wind, crossed the Termination Shock when the magnetic field suddenly increased by a factor of roughly 3 in mid December and remained high afterwards when Voyager 1 had entered the shock transition and the heliosheath. At the time of writing it is still in the heliosheath at high magnetic field levels heading for the heliopause.

Voyager 2 in the southern heliosphere crossed the Termination Shock with working instrumentation. Magnetic field and plasma measurements across the Termination Shock were published by \cite{Burlaga2008} and \cite{Richardson2008} respectively. Figure \ref{chapTS-fig-BPL} shows three minutes of the Termination Shock crossing TS-3 \citep{Burlaga2008} in 48 s time resolution around the Termination Shock ramp. The thin horizontal line in the uppser panel is the value of the solar wind magnetic field prior to the crossing. Shown here are the shock foot, ramp and overshoot region. Afterwards the magnetic field dropped to the heliosheath value which is a factor of $B_2/B_1\approx 1.7$ higher than the solar wind magnetic field identifying the shock as a magnetically relatively weak compression. (Note that the overshoot is roughly a factor 5 higher, which is due to currents flowing in the overshoot region!) 

The solar wind magnetic field was at a Parker spiral direction of $\sim 270^\circ$ azimuthal angle. Entering the foot and shock ramp this angle deviates up to $\sim 30^\circ$ from this direction. The inclination angle in the solar wind fluctuated around $-60^\circ$, in the foot it is more around $30^\circ$ showing oscillation between $0^\circ$ and $\sim 50^\circ$. similar oscillations occur in the ramp. The angular trace shows that the shock returns already in the overshoot to the spiral direction. It is the inclination which catches the change in shock direction becoming close to $0^\circ$ in the overshoot and heliosheath. Altogether the changes in direction are relatively small. The shock crossing TS-3 is hence about quasi-perpendicular and is also supercritical as it develops a well pronounced foot. The Mach number was about ${\cal M}\approx 5$.

There are strong and irregular fluctuations in the magnitude of the foot magnetic field which must be related to the reflected gyrating particle distribution which carries the foot currents and excites electromagnetic waves. The fluctuations in the ramp are very large amplitude. From the full particle simulations described in Chapter \ref{chap4-quasiperpendicular shocks} we know that these fluctuations are directly related to the trapping of ions in the ramp region with the magnetic maxima coinciding with the vertexes in the ion velocity distribution, as shown in Figure \ref{chap4-fig-final}. They are thus indications of the ion dynamics, formation of vortices, ion heating and acceleration. In the overshoot these fluctuations become shorter wavelength which is also known from the simulation results and is related to the smearing out of the ion phase space vortices and the strong ion heating in the overshoot region. 

When comparing these observations with the plasma wave observations by \cite{Gurnett2008} he interesting observation is made that the entire broadband noise spectrum observed in the plasma waves is restricted to the time interval 0010:45-0013 UT which is the transition from the ramp into the overshoot. This is the region where the overshoot currents flow. This is not unexpected and is in agreement with the {\footnotesize AMPTE IRM} observations of the broadband noise spectra. From the numerical simulations \citep{Matsukiyo2003,Matsukiyo2006a,Matsukiyo2006b} it is also known that in this region the ion flow is stopped and is most strongly heated in a quasi-perpendicular shock with the trapped ion vortices loosing identity. 

As noted above, the magnetic field observations during the Voyager 2 passage across the Termination Shock identified five Termination Shock crossings within roughly 12 hours starting with an inbound shock crossing and ending with an inbound crossing. This is possible only for the Termination Shock being highly dynamical and oscillating back an forth which can have different reasons, the most probable are that the shock is unstable with respect to waves that travel along its extended surface causing ripples on the shock surface. Such waves can be assumed to travel at a velocity that is of the order of the local Alfv\'en speed, which is of the order of $V_A\sim 150$ km/s. The times between the crossing TS-1 to TS-5 were 3 hors, 4 hours, and 2 hours, respectively. Thus the average wavelength of those ripples or surface waves will be of the order of $\sim 30000$ km. Since the ion inertial length is $\lambda_i\sim 6000$ km, this is not more than a surface wave wavelength of $\lambda_{sfw}\sim 5\lambda_i$. 

The plasma data \citep{Richardson2008} of the Termination Shock crossings did, in principle, not provide much surprises. They confirm that the shock causes a steep decrease in the flow velocity and a comparable increase in plasma density which is in agreement with the notion of the Termination Shock being a quasi-perpendicular shock. Except for the determination of the bulk plasma parameters the main conclusion is that the solar wind plasma is not heated strongly when crossing the Termination Shock. The bulk of the energy lost by the solar wind stream goes into the pickup ions. Only when taking into account this effect, \cite{Richardson2008} were able to satisfy the necessary shock condition that the Termination Shock decelerates the solar wind flow to Mach  number ${\cal M}\lesssim 1$. This is what could have been expected from the simulations \citep{Kucharek1995,Scholer1999a} however, which showed that pickup ions benefit most in energy by being reflected from the shock. They can be accelerated up to energies several ten times the upstream flow energy.  While this effect is strongest at quasi-parallel shocks, there is some discrepancy left between the interpretation of the observations and the predictions from the simulations. Possibly quasi-perpendicular shocks are also good reflectors of pick-up ions.  

\subsection{Energetic particles}
\noindent As we have noted the greatest interest in the Termination Shock is from the side of the acceleration of ions (and possibly also electrons) to high energies and into the range of Cosmic Ray energies. Theory has predicted that the Termination Shock would serve as the source region of the Anomalous Cosmic Ray component detected in the heliosphere. This component is clearly accelerated only in the heliosphere or in its boundary region and consists of high energy interstellar ions which are produced when entering the heliosphere. Anomalous Cosmic Rays are thus a subset of the interstellar pickup ions in the heliosphere. The Voyager 1 and Voyager 2 Termination Shock crossings have therefore been looked forward with the greatest expectations. Would the Termination Shock be identified as the ultimate source of the Anomalous Cosmic Ray component. 
\begin{figure}[t!]
\centerline{\includegraphics[width=0.8\textwidth,clip=]{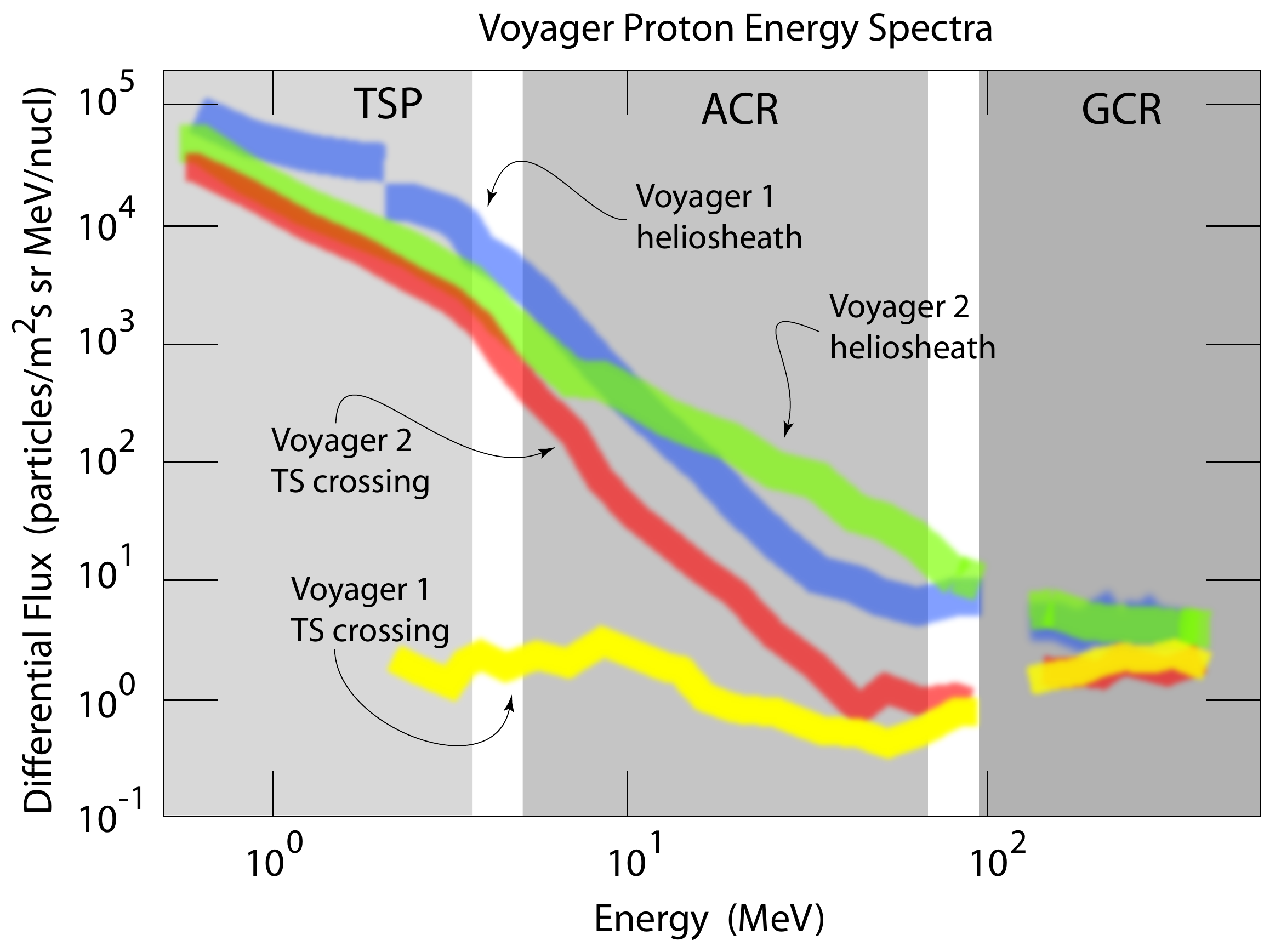}}
\caption[V2]
{\footnotesize The Voyager high energy proton spectra at Termination Shock crossing and in the heliosheath \citep[data taken from][]{Stone2008}. Shown are the proton differential energy fluxes per nucleon (in the case of the protons this is per particle) as function of energy. The width of the lines corresponds about to the error or the measurement. The energy range is divided into the three different populations: Termination Shock protons $< (3-5)$ MeV, Anomalous Cosmic Rays $(6-60)$ MeV, Galactic Cosmic Rays $>90$ MeV. }\label{chapTS-fig-Stone}
\end{figure}

As noted above, the plasma observations at the low energy end of the particle distribution have clearly shown that it is not the plasma that picks up the bulk energy loss of the flow across the Termination Shock. Neither is the plasma heated sufficiently, nor is it decelerated sufficiently. The deceleration could be accounted for by including the pickup ion component \citep{Richardson2008}. In this section we now analyse the Voyager observations of both, the energetic particle observations \citep{Decker2005,Decker2008} and the measurements of the Anomalous Cosmic Ray component \citep{Stone2005,Stone2008}. 

Figure \ref{chapTS-fig-V1gur} in the lower panel on its right shows the big surprise provided by the energetic particle observations during the first Termination Shock crossing by Voyager 1. All expectations culminated in finding the maximum in the energetic ions (and possibly even electrons) right at the Termination Shock crossing. Surprisingly this was not the case. Instead, about 50 days before the shock crossing the energetic paritcles $>70$ MeV/nucleon dropped gradually to a minimum that relaxed during approach to the shock. However, the fluxes of energetic particles increased almost continuously with progressing distance of the Voyager 1 spacecraft from the Termination Shock when moving into the heliosheath. This observational result was not only most surprising, it also was taken as killing the assumption that the Anomalous Cosmic Ray component is caused at the Termination Shock by a shock surfing/first-order Fermi acceleration process acting on the interstellar pickup ions in the outer heliosphere. Indeed, when the maximum of the Anomalous Cosmic Ray flux is not found at the location of the Termination Shock but much farther out in the heliosheath or possibly even at the heliopause or in the ion wall transition region or the interstellar bow shock it is difficult to accept that the Termination Shock accelerates these particles to the high energies they possess in the inner heliosphere.

\subsubsection{Galactic Cosmic Rays}
\noindent Now, after Voyager 2 has crossed the Termination Shock the particle spectra at the two spacecraft can be compared. This has been done in Figure \ref{chapTS-fig-Stone} which shows the differential energy fluxes of the energetic protons as function of energy per nucleon. The plot is divided into the three particle components Termination Shock protons TSP of energy $<(3-5)$ MeV, Anomalous Cosmic Rays in the range $(6-60)$ MeV, and Galactic Cosmic Rays $>90$ MeV. 

There is quite a number of interesting results in this figure. The first is that the GCR spectra for the two Termination Shock crossings at the different locations in the heliosphere do about agree. The Voyager 1 crossing being farther out saw a slightly more intense GCR flux. However the difference is not large suggesting that at the two locations the heliospheric / heliosheath screening effect on the GCR component is comparable. In other words, even though the crossing of the Termination Shock at Voyager 2 was roughly 10 AU farther inside the heliosphere than the Voyager 1 crossing, the boundary of the heliosphere cannot have been much farther out than at Voyager 1. This agrees with the claim that the entire heliosphere is deformed being smaller in its southern than in its northern part. Since the Lyman $\alpha$ observations of \cite{Lallement2005} it is known that this deformation is the result of the presence of the comparably strong interstellar magnetic field that penetrates the LISM in the Local Interstellar Cloud (see Figure \ref{chapTS-fig-TS}). With further progress into the heliosheath the GCR intensities on both spacecraft have increased at the same rate and are practically (within the error of observation) equal, again reflecting the heliospheric filter effect which cause the positive radial gradient in the GCR flux. In principle from these observations the heliospheric cosmic ray diffusion coefficient could be determined.

\subsubsection{Termination Shock Particles}
\noindent These particles are essentially  the pickup ions that are generated in the charge exchange process between the interstellar gas that penetrates the heliosphere plus a small fraction of upstream protons from the solar wind flow that have been reflected and accelerated at the Termination Shock. The acceleration of pickup ins proceeds in two steps. first they become accelerated in the upstream convection electric field ${\bf E}_u=-{\bf V}_u\times{\bf B}_u$ until they merge into the main flow. In this process they perform cycloid orbits in the plane perpendicular to the magnetic field and reach velocities $v=2V_u$ twice as high as the streaming velocity or energies per nucleon ${\cal E}_{pu}/{\rm nucleon}=4{\cal E}_u$ four times the upstream streaming energy. In a second step then they are reflected from the shock and further accelerated. Remember that pickup ions are much better reflected from a shock than main flow particles and should thus become much more accelerated than upstream protons. We should, however, keep in mind that this is the expectation from essentially one-dimensional numerical hybrid, test particle and also PIC simulations of quasi-perpendicular respectively quasi-parallel shocks, which is to be proved by observations. 

It is interesting to compare the conditions at 1 AU and at the location of the Termination Shock at 80-100 AU. Since the nominal solar wind speed $V_u\sim 500$ km/s is assumed to be constant throughout the heliosphere, the motional electric field $E_u$ decreases with distance as the interplanetary magnetic field. At 1 AU its value is $E_u\sim 2.5$ mV/m. At the location of the Termination Shock it is  roughly two orders of magnitude less in the interval $0.025<E_u<0.03$ mV/m. One might thus suspect that its role is diminished. However, what counts are the scales. Because of the weaker magnetic field in the outer heliosphere the gyroradius and thus the radius of the pickup particle cycloid orbit is two orders of magnitude larger such that the pickup ion becomes accelerated at a hundred time slower rate but reaches the same energy as at 1 AU. On the other hand, the energy of a particle that is reflected from the shock, experiences the upstream motional electric field and becomes accelerated along the shock surface depends on the scale tangential to the shock surface. At 1 AU this scale is typically of the order of an Earth radius 1 R$_{\rm E}$ for acceleration at the Earth's bow shock. The shock reflected particle thus gains an energy $\Delta\,{\cal E}_{\rm BS}\sim 16$ keV in one reflection. The typical scale in the outer heliosphere is 1 AU, however.  Hence, the typical energy gain of  a shock reflected particle at the Termination Shock is in the range $3.75<\Delta\,{\cal E}_{\rm TS}<4$ MeV, which shows the different order of magnitude between the effects of a small planetary bow shock and the huge heliospheric Termination Shock.

Energy spectra of the Termination shock particles (TSP) are shown on the left in Figure \ref{chapTS-fig-Stone}. Voyager 1 detected a low and about constant pickup ion flux of the order of $\sim 2$ particles/(m$^2$\,s\,sr\,MeV/nucleon) in the TSP range. This contrasts with the 3-4 orders of magnitude higher TSP fluxes detected in the southern heliosphere at the Voyager 2 Termination Shock crossing. As expected, in the heliosheath the Voyager 2 fluxes increase weakly gradually becoming comparable to the Voyager 1 heliosheath fluxes in the TSP range. These have gradually increased by 3-4 orders of magnitude when Voyager moved into the heliosheath a fact that is not understood yet. Note that the latter three spectra break at about $\sim 3$ MeV, suggesting that the TSP proton energy range ends here, while the ACR proton energy range starts from here up. An indication of this break is also seen in the Voyager 1 shock crossing curve where the values above 3-4 MeV can be interpreted as already being ACR particles.

\subsubsection{The Anomalous Cosmic Ray component}
\noindent The measurement of the ACR component is shown in the central part of Figure \ref{chapTS-fig-Stone}. Voyager 1 observed very low ACR fluxes when crossing the Termination Shock. In fact, when interpreting the spectral break at roughly (3-5) MeV as the transition from pickup ions to ACR then the maximum around $\sim$10 MeV is the increase due to the residual ACR seen by Voyager 1 at the Termination Shock crossing. These ACR cannot have come from the TSP pickup ions at the Voyager 1 position, however. Their origin must be from remote. The spectral increase above $\sim$50 MeV is the low energy extension of the Galactic Cosmic Ray spectrum. The latter is also seen in the Voyager 2 Termination Shock crossing. 

The Voyager 2 Termination Shock crossing finds a much higher ACR flux which can well be interpreted as being the high energy tail of the TSP particle spectrum starting at the spectral break at $\sim$(3-5) MeV and merging with the GCR spectrum at $\sim$50 MeV. In the heliosheath Voyager 1 observes the same ACR spectrum however a factor of 2-3 more intense thus suggesting that Voyager 1 is approaching the source of the ACR in the heliosheath while, obviously, the Termination Shock crossing of Voyager 2 is still not in the ACR acceleration region. But the entrance of Voyager 2 into the heliosheath brings a new surprise. The ACR spectrum follows only in its low energy part at $\sim$(3-8) MeV the Voyager 2 Termination Shock and Voyager 1 heliosheath ACR spectrum. From $\sim$8 MeV on to roughly $\sim$100 MeV it substantially exceeds the Voyager 1 heliosheath fluxes until it merges into the GCR spectrum which takes over at the higher energies. 
\begin{figure}[t!]
\centerline{\includegraphics[width=1.1\textwidth,clip=]{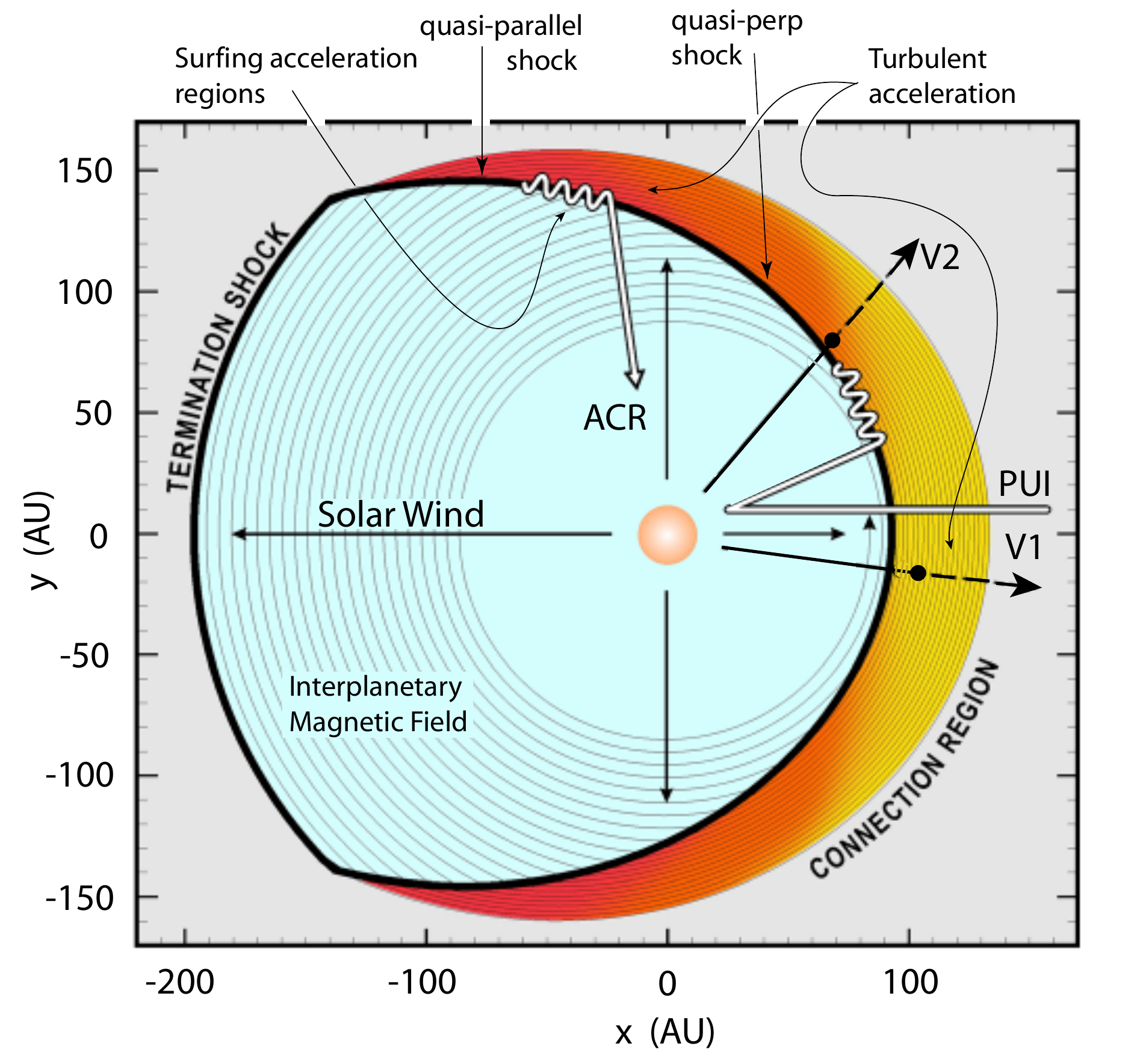}}\vspace{-0.2cm}
\caption[V2]
{\footnotesize The ecliptic projection of the orbits of the Voyager 1 and Voyager 2 spacecraft, a blunt nose model of the Termination Shock, and the interplanetary magnetic field spiral \citep[basic drawing taken from][]{McComas2006}. In this model the spiral interplanetary magnetic field lines cross the Termination Shock at the flanks of the Termination Shock where the shock is effectively quasi-parallel, suggesting that the Anomalous Cosmic Ray component is accelerated at the TS flanks and are thus not seen in the Voyager 1 TS crossing at close to heliospheric nose location where the Termination Shock is quasi-perpendicular. It should also see lower fluxes of accelerated pickup ions PUI. Voyager 2 should observe higher accelerated ACR fluxes. According to this picture Voyager 2 would still be in the quasi-perpendicular shock. Two types of acceleration shown are surfing of particles along the shock and turbulent acceleration in the extended heliosheath. In addition particles could be accelerated bouncing between the footpoints of one spiral field line on the Termination Shock through the turbulent heliosheath. }\vspace{-0.4cm}\label{chapTS-fig-Comas}
\end{figure}

Comparing the spectral slopes we find that the Voyager 2 Termination Shock crossing and the Voyager 1 heliosheath ACR spectra (after the spectral break) have approximate power law shapes with power $\alpha\sim -3.2$, while the Voyager 2 heliosheath spectrum has power law $\alpha\sim -1.5$. Voyager 2 thus is in a region where the acceleration process of the Anomalous Cosmic Ray component is most active and probably caused by turbulent acceleration \citep{Fisk1974,Fisk2006a,Fisk2006}. Thermodynamic arguments for turbulent power laws have been given by \cite{Fisk2007} and \cite{Treumann2008}. In this respect it is of interest that the Voyager 1 TSP proton fluxes below $\lesssim$7 MeV have since remained about constant indicating that Voyager 1 has entered the heliosheath region where the flux has reached maximum and remains to be homogeneous while the ACR fluxes in the range from 12 MeV/nucleon to 22 MeV/nucleon on both spacecraft, Voyager 1 and Voyager 2, continue to increase in intensity at about similar rates {\citep{Stone2008}. At higher energies $\gtrsim$60 MeV/nucleon the increases and fluxes are identical and slow indicating the gradual contribution of the GCR component with approaching heliopause.

The tantalising question that comes up when looking at these measurements is what the role of the Termination shock is in the acceleration of the pickup ions and generation of the Anomalous Cosmic Ray component in the heliosphere. The observations contradict the simple idea of shock acceleration by multiple reflection in which case the maximum of ACR intensity should be found at and around the location of the Termination Shock. Since this is not the case for the Voyager 1 passage, the Voyager 1 orbit did not cross the region of particle acceleration at the shock. On the other hand, the Voyager 2 shock crossing would not be in disagreement with the shock acceleration role would not the fluxes at the ACR energies for Voyager 2 as well increase farther out in the heliosheath causing the spectrum to be extraordinarily (though not impermissibly) flat in the ACR energy range. Obviously the example of the Termination shock shows that the mechanism of shock acceleration is not very well understood yet.  

A possible though simple geometric interpretation has been put forward by \cite{McComas2006}. This model is shown in Figure \ref{chapTS-fig-Comas}. It proposes a blunt Termination Shock with the spiral magnetic field lines of the interplanetary field cutting through the Termination shock at the flanks of the heliosphere. Similar geometries had already been used by \cite{Gurnett2005} in their explanation of the observation of upstream Langmuir waves through magnetic connectivity to the Termination Shock. The assumption of \cite{McComas2006} that has been repeated and elaborated on in follow up papers \citep{Schwadron2007,Schwadron2008} is that the ACR particles are accelerated along the flanks of the Termination Shock where they have sufficient time to surf along the Termination Shock. However, when flowing away from the shock along the field lines they will be found at the heliosphere nose position to occur farther out in the heliosheath. Thus the Voyager 1 observations are lacking these high fluxes which is in agreement with just seeing a slight enhancement around 10 MeV at Voyager 1 during the Termination Shock crossing and the spectral increase of ACR at Voyager 1 in the deep heliosheath. The 10 MeV increase in this model is just the particles of the pickup ion distribution that are accelerated close to Voyager 1 at the shock . The prediction for Voyager 2 to observe much higher fluxes is also in agreement with observation and does well apply to the Voyager 2 crossing of the Termination Shock and the power law tail extending the pickup ion spectrum into the ACR range. 

The two observations of the very low energetic pickup ions at the Voyager 1 crossing and the enormous increase in the ACR spectra above 10 MeV far over what Voyager 1 observed is not really explained by this purely geometrical model, however. Obviously there are much less pickup protons at the Voyager 1 location. Close to the nose 10 MeV pickup ions must have been accelerated from roughly 4 keV/nucleon initial energy by a factor of $2\times 10^4$ implying considerably more than one reflection from the Termination Shock, which implies a large area on the Termination Shock surface being involved in ion reflection. On the other hand, the flat spectrum of the ACR seen by Voyager 2 in the heliosheath suggests the strong involvement of heliosheath turbulence in the acceleration process. This is very similar to acceleration of energetic particles in ordinary solar wind turbulence. The heliosheath seems to host rather high turbulence levels that are capable of accelerating ACR. One would thus tentatively conclude that the Termination Shock is not the main source of the Anomalous Cosmic Rays; the Termination Shock is, however, necessary for ACR in order to produce the high turbulence in the heliosheath as well as the seed population of pre-accelerated pickup ions which are accelerated further outside the Termination Shock by the heliosheath turbulence while still being confined to the heliosheath magnetic field. Voyager 2 being at the foot points of the average heliosheath magnetic field at the flank of the Termination Shock does observe the accelerated ACR particles while Voyager 1 is about to enter the heliosheath sufficiently deep for observing them. The \cite{McComas2006} model and these conclusions are all very speculative and have to be confirmed by future continuing observations provided by the two spacecraft. 

\section{Conclusions}
\noindent The Termination Shock of the heliosphere is the remotest shock in the entire universe that has been subject to {\it in situ} observation by two spacecraft, Voyager 1 and Voyager 2. The next shock that is expected to be crossed by them, the heliospheric Bow Shock in the interstellar medium lies decades ahead. The Voyager crossings have not simply confirmed the theoretical predictions about the structure and nature of the largest -- in the sense of dimensions -- shock in the heliosphere. They have also added new problems and new knowledge which will form the base of future thought, modelling and investigation based on the available data. 

It has been believed that the Termination Shock is a completely perpendicular or quasi-perpendicular shock. The observations have confirmed that -- in the average -- the Termination Shock is indeed quasi-perpendicular in the regions crossed by the two spacecraft. On the other hand it has also been found that it is variable on various time scales. Over large distances of several AU it has been occasionally connected to the Voyager spacecraft long before they ultimately entered and crossed the shock. This connection was identified from typical shock reflected particle fluxes, both ions and electrons, in sunward direction and from the observation of the Langmuir waves that are excited by electron beams along the spiral interplanetary magnetic field. One of the great surprises is the absence of any radio radiation that is emitted from the Termination Shock which for other shocks in the solar system and heliosphere is one of the main shock identifiers from remote. The conditions why the Termination Shock is not radiating may lie in the Termination Shock environment. Possibly the particles are too cool. But no firm explanation has been given yet. 

In plasma waves the Termination Shock has turned out to behave like a strong collisionless supercritical shock possessing a narrow $\sim 1\lambda_i$ wide region of very large amplitude broadband plasma wave turbulence covering the shock ramp and overshoot transition the source of which must be sought in the electron dynamics inside the shock. This turbulence will sign responsible for particle trapping, surfing and heating, i.e. for  transport and acceleration processes in the shock. In the magnetic field the Termination Shock shows the typical signs of an extended foot, ramp and overshoot, signatures of ion trapping and formation of ion phase space vortices with maximum magnetic field strength at their vertexes. 

Our current knowledge of shock dynamics lets us conclude that the Termination Shock is capable of reflecting ions which are accelerated in the foot, and the shock will be subject to reformation that is driven by the reflected gyrating ion fluxes and by current instabilities like the modified two-stream instability. These foot currents and the quasi-periodic reformation modulate the shock along its surface, causing ripples and variations in the shock normal direction, sometimes providing it a quasi-parallel character which explains the occasional remote magnetic connectivity. However, the Termination Shock seems not to be such a strong shock as had been expected. Though being supercritical, it has compression ratio $\lesssim 2$ which might be an indication of mediation of the shock by the high density of pickup ions that are present in its environment.    

A peculiarity of the Termination Shock is that it is embedded into the spiral solar wind magnetic field which is divided into sector structures roughly $0.3$ AU apart in radial direction. These field lines may cut the Termination Shock at two positions on the flanks forming extended loops in the heliosphere with foot points on the Termination Shock. This kind of geometry may affect the property of the Termination Shock to accelerate pickup ions to the high energies of ACR when confining the shock surfing particles downstream in the loop configuration. Fluctuations of the foot points on the Termination Shock of these downstream spiral loops which can be caused by shock surface waves and oscillations of the Termination Shock and breathing of the heliosheath magnetic field in combination with the heliosheath turbulence may be responsible for heating and acceleration. 

We may state in summary that the Termination Shock acceleration mechanism of charged particles in the energy range of the Anomalous Cosmic Rays is not yet understood, and the Termination Shock has not contributed yet sufficiently to its ultimately clarification, even though this clarification was expected to be achieved from the Voyager crossings. At the contrary, it seems that acceleration of particles into ACR is the result of a complicated interplay between the Termination Shock, the local geometry of the spiral interplanetary magnetic field at its contact with the Termination Shock, and the heliosheath turbulence. Acceleration at the Termination Shock is not a process that is homogeneously distributed over the shock surface. It is regional, apparently being restricted to certain regions on the Termination Shock surface, and is otherwise affected by the conditions and the state of turbulence in the heliosheath.

%\vspace{-0.4cm}
%{\acknowledgements This study arose from an ISSI Working Group cooperation on Collisionless Shocks in the Heliosphere. 
%}

\end{document}